\documentclass[oldversion]{aa}
\usepackage{graphicx}
\begin{document}

\title{Diffuse interstellar bands in RAVE Survey spectra\thanks{Table 1 is available electronic only}}

   \author{U.~Munari\inst{1}, 
           L.~Tomasella\inst{1},
           M.~Fiorucci\inst{1},
           O.~Bienaym\'e\inst{2},         
           J.~Binney\inst{3},            
           J.~Bland-Hawthorn\inst{4},     
           C.~Boeche\inst{5}
           R.~Campbell\inst{5,6},
           K.C.~Freeman\inst{7},            
           B.~Gibson\inst{8},
           G.~Gilmore\inst{9},           
           E.K.~Grebel\inst{10},            
           A.~Helmi\inst{11},             
           J.F.~Navarro\inst{12},           
           Q.A.~Parker\inst{6}  
           G.M.~Seabroke\inst{9,13}          
           A.~Siebert\inst{2,5},          
           A.~Siviero\inst{1},
           M.~Steinmetz\inst{5},          
           F.G.~Watson\inst{14},              
           M.~Williams\inst{5,7}
           R.F.G.~Wyse\inst{15},               
           T.~Zwitter\inst{16}}             

   \offprints{ulisse.munari@oapd.inaf.it}

  \institute{INAF Osservatorio Astronomico di Padova, Asiago, Italy
      \and Observatoire de Strasbourg, Strasbourg, France
      \and Rudolf Pierls Center for Theoretical Physics, University of Oxford, UK
      \and Institute of Astronomy, School of Physics, University of Sydney, Australia
      \and Astrophysikalisches Institut Potsdam, Potsdam, Germany
      \and Macquarie University, Sidney, Australia 
      \and RSAA Mount Stromlo Observatory, Camberra, Australia
      \and University of Central Lancashire, Preston, UK
      \and Institute of Astronomy, University of Cambridge, UK
      \and Astronomisches Rechen-Institut, Center for Astronomy of the University of Heidelberg, Heidelberg, Germany
      \and University of Groningen, Groningen, The Netherlands
      \and University of Victoria, Victoria, Canada
      \and e2v Centre for Electronic Imaging, School of Engineering and Design, Brunel University, Uxbridge, UK
      \and Anglo Australian Observatory, Sydney, Australia
      \and Johns Hopkins University, Baltimore, Maryland, USA
      \and Faculty of Mathematics and Physics, University of Ljubljana, Ljubljana, Slovenia
              }

   \date{Received YYY ZZ, 2005; accepted YYY ZZ, XXXX}

     \abstract{We have used spectra of hot stars from the RAVE Survey in
               order to investigate the visibility and properties of five
               diffuse interstellar bands previously reported in the
               literature.  The RAVE spectroscopic survey for Galactic
               structure and kinematics records CCD spectra covering the
               8400-8800~\AA\ wavelength region at 7500 resolving power. The
               spectra are obtained with the UK Schmidt at the AAO, equipped with the 6dF
               multi-fiber positioner. The DIB at 8620.4~\AA\ is by far the
               strongest and cleanest of all DIBs occurring within the RAVE
               wavelength range, with no interference by underlying
               absorption stellar lines in hot stars. It correlates so
               tightly with reddening that it turns out to be a reliable
               tool to measure it, following the relation $E_{B-V} = 2.72
               (\pm 0.03)\times E.W.$(\AA), valid throughout the general
               interstellar medium of our Galaxy. The presence of a DIB at
               8648~\AA\ is confirmed. Its intensity appears unrelated to
               reddening, in agreement with scanty and preliminary reports
               available in the literature, and its measurability is
               strongly compromised by severe blending with underlying
               stellar HeI doublet at 8649~\AA. The two weak DIBS at 8531
               and 8572~\AA\ do not appear real and should actually be
               blends of underlying stellar lines. The very weak DIB at
               8439~\AA\ cannot be resolved within the profile of the much
               stronger underlying hydrogen Paschen~18 stellar line.

             }

   \authorrunning{U.Munari et al.}
   \titlerunning{DIBs in RAVE Survey spectra}

   \maketitle

\section{Introduction}

  \begin{figure*}
     \centering
     \includegraphics[height=18.0cm,angle=270]{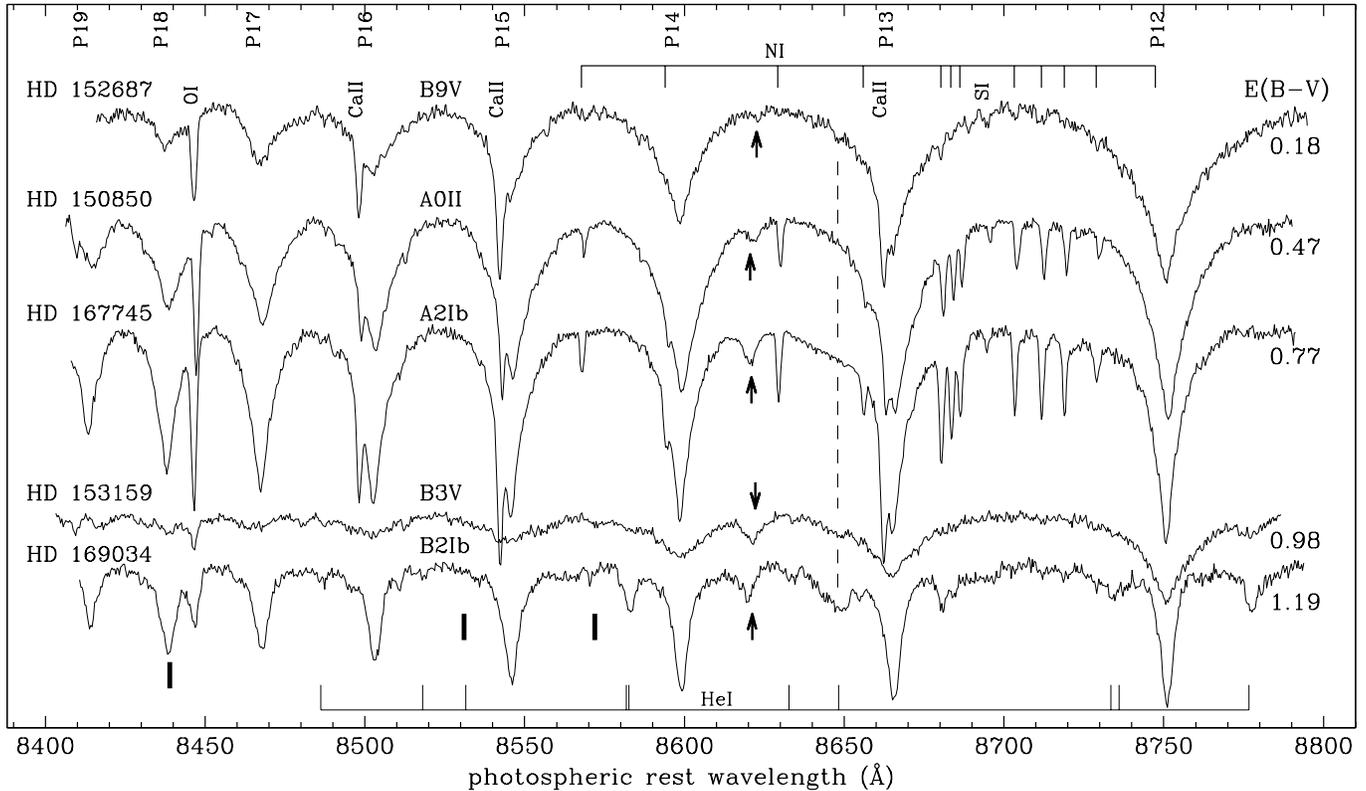}
     \caption{A sample of RAVE spectra of early type HD stars ordered
     according to reddening. The strongest stellar lines are identified. The
     arrows point to DIB 8620, the thick
     vertical marks to DIBs 8439, 8531, 8572, and the dashed line to
     DIB 8648~\AA.}
     \label{fig1}
  \end{figure*}

Diffuse interstellar bands (DIBs) were first discovered by Heger (1922) as
stable lines that did not follow the orbital motion seen in spectroscopic
binaries. The first systematic studies began only with Merrill (1934, 1936),
who noted similarities (occurrence, intensity and velocity) and differences
(far wider and with diffuse edges) with respect to atomic interstellar
lines. A few years later the first interstellar molecule (CH) was identified
by means of absorption lines at 4300.3~\AA\ by Swings and Rosenfeld (1937).
Since then, much progress has been recorded in understanding the absorption
spectra of interstellar ions and molecules, while the origin of DIBs is
still mysterious after almost a century after their discovery (Sarre 2006).
The census of major DIBs in the optical region seems quite complete now, at
least for those away from strong telluric absorption bands and stellar
lines.  A compilation maintained by Jenniskens (2007) lists 281 DIBs over
the wavelength range from 3980 to 9632~\AA. DIBs are also observed in
external galaxies (e.g. in the SMC by Cox et al. 2007b; in the LMC by Cox et
al. 2006; in NGC 1448 by Sollerman et al. 2005; in M31 by Cordiner et al.
2008), in starbust complexes (Heckman and Lehnert 2000) and damped
Ly-$\alpha$ systems (Jukkarinen et al. 2004, York et al. 2006, Ellison et
al. 2008)

The carriers of DIBs are still unknown. Some DIBs tend to show an
appreciable correlation with reddening, even if others do not (e.g. Sanner
et al. 1978, Krelowski et al. 1999), and this led to the hypothesis that
they were produced on or in the interstellar dust grains. However,
lack of polarization in DIB profiles (Cox et al. 2007a) argues against, and
complex carbon-bearing molecules are generally considered as viable carriers
(e.g. Fulara and Krelowski 2000), with fullerenes being subject of intensive
laboratory studies (e.g. Herbig 1995, Leach 1995, Iglesias-Groth
2007). PAH (polycyclic aromatic hydrocarbons) have been frequently
considered as promising carriers (van der Zwet and Allamandola 1985, Leger
and Dhendecourt 1985), in particular for their success in explaining the
unidentified infrared emission bands (UIR; Sarre et al. 1995). Laboratory
studies suggest that only ionized and not neutral PAH could be viable
carriers (e.g. Ruiterkamp et al. 2002, Halasinski et al. 2005), and this
would agree with the observed insensitivity of DIB intensity to ambient
electron density (Gnaciski et al. 2007). A search for correlations between
different DIBs has been carried out as a criterion to guide possible
identification (a strict correlation may imply a common carrier, whereas a
lack of correlation indicates that different species are involved), but the
degree of correlation cover the whole interval from good to very poor (e.g.
Moutou et al. 1999). Finally, Holmlid (2008) has recently proposed a
radically different mechanism for the formation of DIBs, namely doubly
excited atoms embedded in the condensed phase named Rydberg matter.

The Radial Velocity Experiment (RAVE) is an ongoing spectroscopic survey of
the whole southern sky at galactic latitudes $|b|$$\geq$25$^\circ$ for stars
in the magnitude interval 9$\leq$$I_{\rm C}$$\leq$12, with spectra recorded
over the 8400-8800~\AA\ range at a resolving power around 7\,500. The 150
fiber positioner 6dF is used at the UK Schmidt telescope of the
Anglo-Australian Observatory. The main scientific driver of the project is
the study of the stellar kinematics and metallicity of galactic populations
away from the galactic plane (e.g. Smith et al. 2007, Seabroke et al. 2008,
Veltz et al. 2008). RAVE begun operations in 2003 and has now reached the
milestones of the first (Steinmetz et al. 2006) and second (Zwitter et al.
2008) data releases, with a current total of $\sim$250\,000 stars already
observed.

In this paper we investigate the detectability, measurability and properties
of the DIBs known to occur within the RAVE wavelength interval. In this
range there is no resonant line from ions sufficiently abundant in the
interstellar medium to produce detectable features in high-resolution ground
spectra. Weak C$_2$ interstellar lines due to the (2,0) band of the Phillips
system are seen longword of hydrogen Paschen~12 and around HeI 8777 stellar
lines (Gredel and Muench 1986) in highly reddened stars, the strongest ones
occurring at 8751.5, 8753.8, 8761.0, 8763.6, 8773.1 and 8780.0~\AA\ (Gredel
et al. 2001).

\section{Target selection}

The flatter is the background stellar continuum, the easier and firmer is
the detection and measure of a DIB. Only hot stars provide continua with a
sufficiently small number of photospheric lines, and all DIB studies in the
literature observed hot stars. Extensive checks with the Castelli and Munari
(2001, hereafter CM01) synthetic spectral atlas, show that spectral types
earlier than A7 have - irrespective of the luminosity class - flat
background continua at the wavelengths of DIB~8620.4, the principal target of
this paper. To be on the safe side by a fairly wide margin, we limited the
target selection to A3 and hotter stars.

DIBs generally increase their strength in pace with reddening, which is
highest at the lowest galactic latitudes. For this reason, we selected for
our analysis only HD stars observed by RAVE at $|b|$$\leq$10$^\circ$ as
part of calibration runs (normal survey observations are carried out at
$|b|$$\geq$25$^\circ$ where the reddening is generally negligible and therefore
DIB signatures undetectable).

We limited the target selection to HD stars because ($i$) for all of them an
accurate and homogeneous spectral classification has been provided by the
Michigan Spectral Survey (Houk and Cowley 1975, Houk 1978, Houk 1982, Houk
and Smith-Moore 1988, Houk and Swift 1999), ($ii$) intrinsic $B-V$ color is
known for all spectral types and luminosity classes (Fitzgerald 1970), and
($iii$) accurate Tycho-2 $B_{\rm T}$ and $V_{\rm T}$ photometry is available
for all targets. Together, they allow to derive a consistent value for the
interstellar reddening affecting each target star.

Two further selection criteria were applied in order to enforce the highest
possible quality of the results: we retained only the spectra ($a$) with
S/N$\geq$50 per pixel on the stellar continuum around 
DIB 8620, and ($b$) that do not show even the most feeble trace
of residual fringing left over by the flat field division. The RAVE Survey
CCD is a thinned, back-illuminated one, and as such it naturally presents
fringing at the very red wavelengths of RAVE Survey observations. Flat
fielding generally provides accurate fringing removal, but in some
cases a weak and residual pattern survives (at a level $\leq$1\% that does
not affect the main RAVE Survey products: radial velocities and atmospheric
parameters of observed stars). 

The final target list after application of all above criteria count 68
targets observed by RAVE during the time interval 28 April 2004 to 21
October 2006, eight of them observed twice. They are listed in Table~1
(available electronic only).

  \begin{figure}
     \centering
     \includegraphics[width=8.8cm]{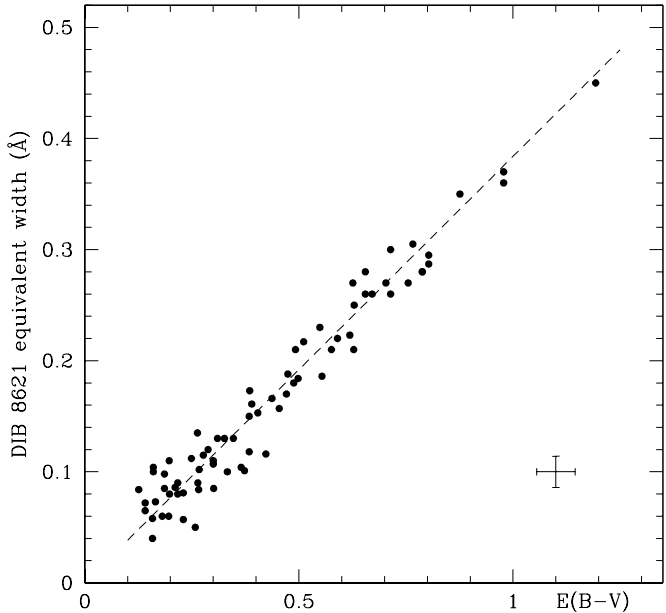}
     \caption{The equivalent width of the diffuse interstellar band
     at 8620.4~\AA\ as function of reddening. The dashed line represents
     Eq.(4) and the error bars are the average uncertainties of plotted points 
     from Eq.(1) and (3).}
     \label{fig2}
  \end{figure}

\section{DIBs over the RAVE wavelength interval}

A DIB at 8439.4~\AA\ was discovered by Galazutdinov et al. (2000). It is very
sharp, and deeply blended with the Paschen~18 photospheric line. It is also
intrinsically quite weak, being detectable only in high resolution and high
S/N spectra of heavily reddened stars. RAVE spectra cannot reveal it.

A sharp DIB at 8530.7~\AA\ was reported by Jenniskens and Desert (1994), but
not confirmed by Galazutdinov et al. (2000). Its wavelength is coincident
with the blend of 8528.967, 8529.025, and 8531.508~\AA\ HeI photospheric
absorption lines (cf CM01), compromising any clear detection. The DIB
equivalent width reported by Jenniskens and Desert (1994) for four early
type stars characterized by $E_{B-V}$ ranging from 0.30 to 1.28, shows no
clear trend with reddening and instead a better correlation with the
intensity of HeI expected from the spectral type of the target star.
Extrapolating Jenniskens and Desert (1994) data for HD~183143 reddened by
$E_{B-V}$=1.28, an equivalent width of 0.065~\AA\ could be expected for this
DIB in the RAVE spectrum of HD~169034 in Fig.~1. This equivalent width is
less than half that of the stellar HeI blend. If the DIB~8530.7 is real,
which we doubt, RAVE spectra cannot disentangle it from the underlying and
overwhelming HeI~8530~\AA\ absorption blend.

Sanner et al. (1978) listed an uncertain DIB at 8572~\AA. It was dismissed
as a photospheric stellar line by Jenniskens and Desert (1994) and was not
detected by Galazutdinov et al. (2000). CM01 atlas shows the presence of a
weak and diffuse blend of many photospheric absorption lines centered at the
same wavelength as the supposed DIBs, confirming the proposed DIB as a
spurious detection.

A moderately strong, broad and complex profile DIB, loosely centered at
8648.3~\AA, was discovered by Sanner et al. (1978). It was later confirmed
by Herbig and Leka (1991), Jenniskens and Desert (1994) and Wallerstein et al. (2007). 
Over the wavelength interval covered by the 8650~DIB there are two strong HeI
photospheric absorption lines at 8648.258 and 8650.811 that appreciably
confuse the picture (cf. CM01). RAVE spectra support the presence of a DIB
at these wavelengths, whose intensity however does not correlate at all with
reddening (as already noted by Sanner et al. 1978, their Fig.~4). In fact,
in the RAVE spectra of Fig.~1, the 8648.3~DIB is clearly present at
$E_{B-V}$=0.18, missing at $E_{B-V}$=0.47, prominent and broader at
$E_{B-V}$=0.77, feeble or absent at $E_{B-V}$=0.98, and again strong at
$E_{B-V}$=1.19. In the most reddened spectrum of Fig.~1 (HD~169034), the
expected intensity of the interfering HeI blend at 8648.258, 8650.811 is
equal to the intensity of the nearby HeI blend at 8581.856, 8582.670, which
lead us to speculate that the equivalent width of the 8648.3~DIB in this
spectrum should be $\sim$0.25~\AA. A very close match is provided by
Wallerstein et al.'s (2007) spectra for the stars HD~169454 and HD~183143 (their
Fig.~1). Given the obvious interference by strong underlying stellar HeI and
the lack of an appreciable correlation with reddening, we will not further
discuss the 8648.3~DIB as seen in RAVE spectra.

\section{The 8620 \AA\ DIB}

The strongest DIB over the RAVE wavelength range appears at 8620.4~\AA. It
was first discovered by Geary (1975), and then confirmed by all later
investigators. Its tight correlation with reddening was discovered and
discussed by Munari (2000, hereafter M00), and later confirmed by 
Wallerstein et al. (2007).

\subsection{DIB measurement}

The equivalent width and heliocentric wavelength of the 8620~DIB has been
obtained on RAVE spectra by integrating - over the wavelength range of the
DIB - the difference between the extrapolated underlying continuum and the
observed spectrum affected by the DIB. The underlying continuum has been
fitting with a 6$^{th}$ order Lagrange polynomial between the Paschen 13 and
14 line centers. The results are reported in Table~1.

Eight of the sixty-eight program stars have a second RAVE spectrum
satisfying the quality selection criteria outlined in Sect. 2. Four of them
have been obtained with similar instrument set-ups (plate and fiber) on
adjacent nights, while the other four were observed with different set-ups one year apart. The
mean difference between the two measurements of the equivalent width in these
eight pairs is:
\begin{equation}
\sigma (E.W.)  ~=~ 0.014~{\rm \AA} 
\end{equation}
which we consider to be representative of the mean accuracy of our DIB measurements.

The intrinsic DIB wavelength is reported as 8620.8~\AA\ by Galazutdinov et
al. (2000) from observations toward a single star (HD~23180), and as
8621.2~\AA\ by Jenniskens and Desert (1994) from observations toward four
stars. M00 gives 8620.4~\AA\ from observations of 37 northern stars in the
general Galactic anti-center direction, after correction for the velocity of
interstellar atomic lines (NaI and KI). The RAVE program stars lies toward the
Bulge and close to the Galactic center (see galactic coordinates in Table~1
and Fig.~3). Adopting the Brand and Blitz (1993) maps for the radial velocity
of interstellar medium, the average velocity of the medium along the lines 
of sight to the program stars is essentially null. Therefore, the mean of heliocentric DIB
wavelength in Table~1 represents also the intrinsic barycentric wavelength,
whose average value is:
\begin{equation}
\lambda (DIB) ~=~ 8620.4 ~(\pm 0.1) ~{\rm \AA}
\end{equation} 

\subsection{Reddening of program stars}

The reddening of the sixty-eight program stars was homogeneously derived
from the Michigan spectral type of the HD stars, the corresponding intrinsic
color $(B$$-$$V$$)_{\rm J}$ from Fitzgerald (1970), and observed the Tycho-2
$(B$$-$$V$$)_{\rm T}$ color ported to the Johnson system via Bessell (2000)
transformations (cf Sect. 2). The reddening derived for the program stars is
listed in Table~1 together with their spectro-photometric parallax derived
adopting absolute $M_V$ magnitudes from the Michigan
Project\footnote{http://www.astro.lsa.umich.edu/users/hdproj/mosaicinfo/
absmag.html}. The individual error sources contributing to the overall reddening
error are: ($i$) the natural color width of a spectral sub-type (on the
average 0.020 mag for O-A3 stars), ($ii$) the uncertainty in the spectral
classification (if taken equivalent to one spectral subclass, it is also
0.020 mag), ($iii$) the error of Tycho-2 $(B$$-$$V$$)_{\rm T}$ color (on
average 0.034 mag), ($iv$) the uncertainty of the color transformation from
Tycho-2 to the Johnson system (unknown, and assumed to amount to a mere 0.010
mag). Considering them as independent quantities and adding them in
quadrature, the mean value of the overall error budget of reddening
determination is:
\begin{equation}
\sigma (E_{B-V}) ~=~ 0.045~{\rm mag} 
\end{equation}

  \begin{figure}
     \centering
     \includegraphics[height=8.8cm,angle=270]{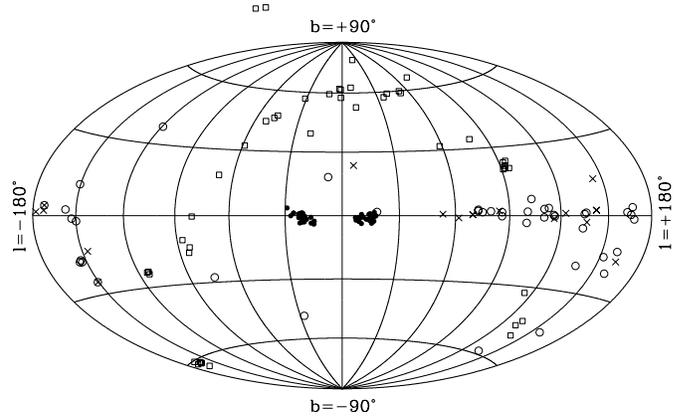}
     \caption{Aitoff projection in galactic coordinates of the program
     stars (filled circles). Also plotted are the stars studied by 
     Munari (2000, open circles), and those observed by
     Jenniskens and Desert (1994), Sanner et al. (1978, crosses)
     and Wallerstein et al. (2007, squares).} 
     \label{fig3}
  \end{figure}

\subsection{A tight relation between reddening and equivalent width}

Figure~2 illustrates the relation between reddening and equivalent width of
DIB~8620.4 for the seventy-six spectra of the sixty-eight program stars of
Table~1. The relation is remarkably straight, with a least square
fitting of:
\begin{equation}
E_{B-V}~=~2.72~(\pm~0.03)~\times~E.W.
\end{equation}
where the E.W. is expressed in \AA. The rms of the points is 0.020~\AA\ in
E.W. and 0.053~mag in $E_{B-V}$, which are only modestly larger than the
typical measurement error in both axes as given by Eq.(1) and (3). This
argues in favor of a very small {\em intrinsic} scatter of the points around
Eq.(4), a remarkable property already preliminary focused upon by M00. The
absence of significant cosmic scatter in the proportionality between
reddening and equivalent width of the DID~8620 has two main implication:
($a$) the DIB carrier has an intimate partnership with the solid phase of
the interstellar medium. A search and study of the polarization across the
DIB profile would be worthwhile, and ($b$) the DIB~8620.4 can now be considered
a viable tool to actually {\em measure} the amount of reddening and not
simply to guess its presence.

The dispersion of the points along the mean relation in Fig.~2 is
slightly larger at lower reddenings: the rms in E.W. of the points with
E.W.$\leq$0.14~\AA\ is 0.022~\AA, and 0.016~\AA\ for E.W.$\geq$0.14~\AA.
While the significance of this small difference is uncertain given the small
number statistics, for sake of discussion it could be argued that some
cosmic scatter - even if marginal - is actually present in the relation
between reddening and intensity of the DIB~8620.4. In fact, a sharper
relation at increasing $E_{B-V}$ could simply mean than one starts to sample
multiple clouds in the line-of-sight and thus any deviations between single
clouds will be somewhat averaged out. Wallerstein et al. (2007) reported that
stars seen through the $\rho$~Oph molecular cloud show a DIB weaker than 
expected.

The proportionality relation found by M00 from 37 northern stars observed at
high resolution and high S/N is $E_{B-V}~=~2.69 (\pm 0.03)~\times~E.W.$,
which is essentially identical to Eq.(4). This is remarkable because the CCD
adopted by M00 for his Echelle observations was a thick, front illuminated
one without any fringing, and therefore ideal to aim for the highest
accuracy. Fitting the data of the four stars observed by Jenniskens and
Desert (1994) in high resolution with a Reticon detector gives
$E_{B-V}~=~2.77 (\pm 0.1) ~\times~E.W.$, and the 10 stars observed by Sanner
et al. (1978) at low resolution again with a Reticon detector provides
$E_{B-V}~=~2.76(\pm 0.06) ~\times~E.W$. Finally, when the reddening of the
target stars is homogeneously computed in the same way as done for this
paper, also the Wallerstein et al. (2007) data support Eq.(4) above 
(G. Wallerstein, private communication).

Our study and data from these other investigations cover about two hundreds
different stars distributed over a great range of distances from the Sun and
over a wide range Galactic longitudes, from the Galactic anti-center mapped
by M00 to the Galactic center in this paper, as illustrated by Fig.~4. These
stars have been observed with quite different techniques and instruments,
and the equivalent width of the DIB measured with different approaches. Yet,
they define one and the same proportional relation. We therefore propose
that Eq.(4) can be safely adopted as a direct way to derive the reddening
caused by the general interstellar medium. 

In general terms, it could be argued that proportionality relation
between reddening and DIB intensity could take the form
$E_{B-V}~=~\alpha(l,b,d)~\times~E.W$, where $\alpha$ is allowed to vary as
function of galactic coordinates and distance, reflecting local
dis-homogeneities in the interstellar medium. The calibration of $\alpha$
requires observations of stars scattered through the whole Galaxy in a
number which is orders of magnitudes larger than currently available. As
noted by M00, only the forthcoming ESA's GAIA mission will be in a position
to provide such a large dataset, both in the form of accurate distances and
spectra covering the DIB~8620.4 in high resolution.

\begin{acknowledgements}
We would like to thank George Wallerstein and Karin Sandstrom for useful
discussions, and the anonymous referee for effective comments. The
spectra here used were obtained as part of the RAVE survey using the UK
Schmidt Telescope operated by the Anglo-Australian Observatory.  The RAVE
project is managed and supported by the Astrophysikalisches Institut
Potsdam, Anglo-Australian Observatory, Australian National University,
University of Basel, University of Cambridge, University of Edinburgh,
University of Heidelberg, Johns Hopkins University, University of Ljubljana,
Macquarie University, University of Oxford, INAF Astronomical Observatory of
Padova, Steward Observatory, Swinburne University, University of Utrecht,
University of Victoria. The RAVE web site is at www.rave-survey.org
\end{acknowledgements}

\clearpage
\newpage
  \begin{table}
     \caption{(available electronic only). Summary of basic information and
     DIB measurements for the program stars. {\em pl} is the RAVE fiber
     plate number (1 or 2), and {\em fiber} is the fiber number of the given
     plate (from 1 to 150). $\lambda_\odot$ is the measured heliocentric
     wavelength of the DIB and {\em e.w.} its equivalent width, both in \AA.
     Photometric magnitudes are in the Johnson system.}
     \centering
     \includegraphics[width=18.0cm,angle=180]{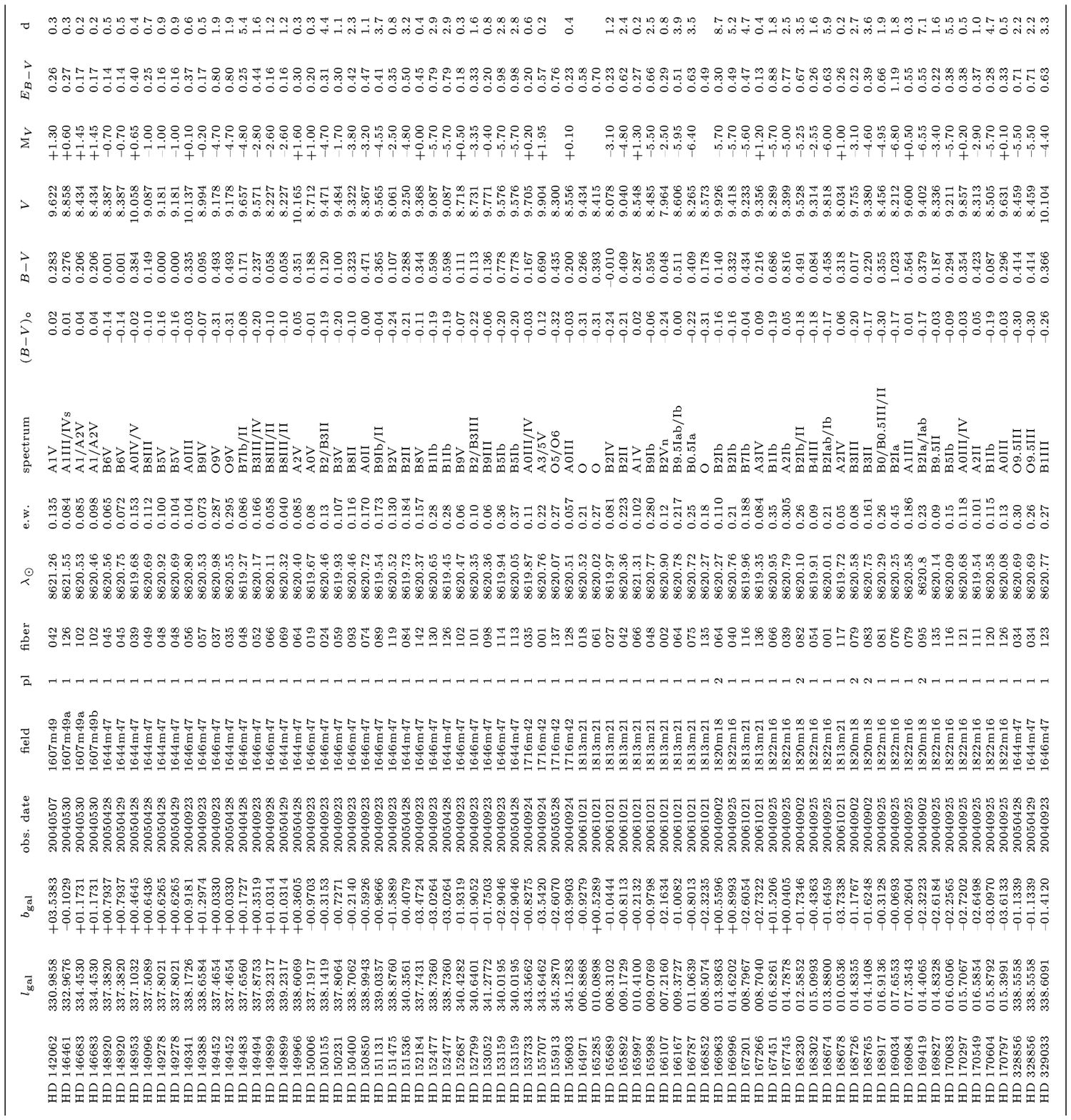}
     \label{tab1}
  \end{table}


\begin{thebibliography}{}
\bibitem[2000]{bessell} Bessell, M.S. 2000, PASP112, 961
\bibitem[1993]Brand, J., Blitz, L. 1993, A\&A 275, 67
\bibitem[2001]{castelli} Castelli, F., Munari, U.  2001, A\&A 366, 1003 (CM01)
\bibitem[2008]{cordiner} Cordiner, M.A. et al. 2008, A\&A 480, L13
\bibitem[2006]{cox06} Cox, N.L.J. et al. 2006, A\&A 447, 991
\bibitem[2007]{cox07a} Cox, N.L.J. et al. 2007a, A\&A 465, 899
\bibitem[2007]{cox07b} Cox, N.L.J. et al. 2007b, A\&A 470, 941
\bibitem[2008]{ellison} Ellison, S.L. et al. 2008, MNRAS 383, L30
\bibitem[1970]{fitzgerald} Fitzgerald, M.P. 1970, A\&A 4, 234 
\bibitem[2000]{fulara} Fulara, J., Krelowski, J. 2000, New Astr. Rev. 44, 581
\bibitem[2000]{gala00} Galazutdinov, G.A. et al. 2000, PASP 112, 648  
\bibitem[1975]{geary} Geary, J.C. 1975, Ph.D. thesis, Univ. Arizona
\bibitem[2007]{gnaciski} Gnacinski, P. et al. 2007, A\&A 469, 201
\bibitem[1986]{gredel86} Gredel, R., Muench, G. 1986, A\&A 154, 336
\bibitem[1986]{gredel01} Gredel, R., Black, J.H., Yan, M. 2001, A\&A 375, 553
\bibitem[2005]{Halasinski} Halasinski, T.M. et al. 2005, ApJ 628, 555
\bibitem[1922]{Heger} Heger, M.L. 1922, LickOB 10, 141
\bibitem[2000]{heckman} Heckman, T.M., Lehnert, M.D.2000, ApJ 537, 690
\bibitem[1991]{herbig91} Herbig, G.H., Leka, K.D. 1991, ApJ, 382, 193
\bibitem[1995]{herbig95} Herbig, G.H. 1995, ARA\&A 33, 19
\bibitem[2008]{holmlid08} Holmlid, L. 2008, MNRAS 384, 764
\bibitem[1975]{houk75} Houk, N., Cowley, A.P. 1975, Michigan Catalogue of Two-Dimensional Spectral Types for the HD Stars, Vol. 1 (Ann Arbor: Univ. Michigan)
\bibitem[1978]{houk78} Houk, N. 1978, Michigan Catalogue of Two-Dimensional Spectral Types for the HD Stars, Vol. 2 (Ann Arbor: Univ. Michigan)
\bibitem[1982]{houk82} Houk, N. 1982, Michigan Catalogue of Two-Dimensional Spectral Types for the HD Stars, Vol. 3 (Ann Arbor: Univ. Michigan)
\bibitem[1988]{houk88} Houk, N., Smith-Moore, M. 1988, Michigan Catalogue of Two-Dimensional Spectral Types for the HD Stars, Vol. 4 (Ann Arbor: Univ. Michigan)
\bibitem[1999]{houk99} Houk, N., Swift, C. 1999, Michigan Catalogue of Two-Dimensional Spectral Types for the HD Stars, Vol. 5 (Ann Arbor: Univ. Michigan)\bibitem[2007]{iglesias} Iglesias-Groth, S. 2007, ApJ 661, L167
\bibitem[1994]{jenniskens94}  Jenniskens, P., Desert, F.X. 1994, A\&AS, 106, 39
\bibitem[1994]{jenniskens07}  Jenniskens, P. 2007, http://leonid.arc.nasa.gov/DIBcatalog.html
\bibitem[2004]{Junkkarinen} Junkkarinen, V.T., Cohen, R.D., Beaver, E.A., Burbidge, E.M., Lyons, R.W., Madejski, G. 2004, ApJ 614, 658
\bibitem[1999]{krelowski} Krelowski, J. et al. 1999, A\&A 347, 235 
\bibitem[1995]{leach} Leach, S. 1995, P\&SS 43, 1153
\bibitem[1985]{leger} Leger, A., Dhendecourt, L. 1985, A\&A 146, 81
\bibitem[1934]{merrill34} Merrill, P.W. 1934, PASP 46, 206
\bibitem[1936]{merrill36} Merrill, P.W. 1936, ApJ 83, 126
\bibitem[1999]{moutou} Moutou, C. et al. 1999, A\&A 351, 680
\bibitem[2000]{munari} Munari, U. 2000, in Molecules in Space and in the Laboratory, I. Porceddu and S. Aiello eds., Ita. Phys. Soc. Conf. Proc. 67, 179 (M00)
\bibitem[2002]{ruiterkamp} Ruiterkamp, R. et al. 2002, A\&A 390, 1153
\bibitem[1978]{sanner} Sanner, F., Snell, R., Vanden Bout, P. 1978, ApJ, 226, 460
\bibitem[2006]{sarre06} Sarre, P.J. 2006, J.Moc.Spc 238, 1
\bibitem[1995]{sarre95} Sarre, P.J., Miles, J.R., Scarrott, S.M. 1995, Science 269, 674
\bibitem[2008]{seabroke} Seabroke, G. M., Gilmore, G., Siebert, A., Bienaymé, O., Binney, J., Bland-Hawthorn, J., Campbell, R., Freeman, K. C. et al. 2008, MNRAS 384, 11
\bibitem[2007]{smith} Smith, Martin C., Ruchti, Gregory R., Helmi, Amina, Wyse, Rosemary F. G., Fulbright, J. P., Freeman, K. C., Navarro, J. F., Seabroke, G. M. et al.  2007, MNRAS 379, 755
\bibitem[2005]{sollerman} Sollerman, J. et al. 2005, A\&A 429, 559
\bibitem[2006]{steinmetz} Steinmetz, M. et al. 2006, AJ 132, 1645
\bibitem[1937]{swings} Swings, P., Rosenfeld, L. 1937, ApJ 86, 483
\bibitem[1985]{allamandola} van der Zwet, G.P., Allamandola, L.J. 1985. A\&A 146, 76
\bibitem[2008]{veltz} Veltz, L., Bienaymé, O., Freeman, K. C., Binney, J., Bland-Hawthorn, J., Gibson, B. K., Gilmore, G., Grebel, E. K., Helmi, A., Munari, U. et al. 2008, A\&A 480, 753
\bibitem[2006]{york} York, B.A. et al. 2006, ApJ 647, L29
\bibitem[2006]{wallerstein} Wallerstein, G., Sandstrom, K., Gredel, R. 2007, PASP 119, 1268
\bibitem[2008]{zwitter} Zwitter, T., Siebert, A., Munari, U., et al. 2008, AJ 136, 421


\end{thebibliography}
\end{document}